\documentstyle[aps,prl,twocolumn,epsf,floats,12pt]{revtex}

\draft
\begin{document}
%%%%%%%%%%%%%%%%%%%%%%%%%%%%%%%%%%%%%%%%%%%%%%%%%%%%%%%%%%%%%%%%%%%%%%
\title{Electron correlations 
       in a partially filled first excited Landau level}
%%%%%%%%%%%%%%%%%%%%%%%%%%%%%%%%%%%%%%%%%%%%%%%%%%%%%%%%%%%%%%%%%%%%%%
\author{
   \underline{Arkadiusz W\'ojs}$^{1,2}$ and
   John J. Quinn$^1$}
\address{\footnotesize\sl
   $^1$Department of Physics, 
       University of Tennessee, Knoxville, Tennessee 37996, USA \\
   $^2$Institute of Physics, 
       Wroclaw University of Technology, Wroclaw 50-370, Poland\\[1em]}
\address{
   \footnotesize\rm\parbox{6.0in}{
   The form of electron correlations in a partially filled degenerate 
   Landau level (LL) is related to the behavior of the anharmonic 
   part of the interaction pseudopotential.
   Unlike in the lowest LL, the pseudopotential in the first excited 
   LL is harmonic at short range.
   As a result, the incompressible states in this LL have different 
   correlations, occur at different filling factors $\nu$, and cannot 
   be described by a composite fermion model.
   The series of Laughlin-correlated states of electron pairs is 
   proposed at $\nu=2+2/(q_2+2)$ with integer $q_2$.
   It includes Moore--Read $\nu={5\over2}$ state and the 
   $\nu={7\over3}$ state.
   Despite coincidence of the values of $\nu$, the latter state 
   has different correlations than Laughlin state of single electrons 
   at $\nu={1\over3}$ and, in finite systems, occurs at a different LL 
   degeneracy (flux).
   }
   \\[-3ex]}
\maketitle

In the absence of another (kinetic) energy scale, correlations 
in a degenerate Landau level (LL) are completely determined by 
the form of electron--electron interaction\cite{prange}.
Depending on the type of these correlations, the series of 
incompressible ground states (GS's) may occur at the specific values 
of the filling factor $\nu$, the elementary excitations of these 
GS's may have specific (quasiparticle) character and, consequently, 
the system (two-dimensional electron gas, 2DEG, in a high magnetic 
field) may exhibit specific optical and transport properties.

For example, it turns out that the short-range character of the 
Coulomb repulsion in the lowest ($n=0$) LL makes the electrons 
maximally avoid those pair eigenstates with the smallest relative 
angular momenta ${\cal R}=1$, 3, 5, \dots
\cite{laughlin,haldane-pseudo,wojs-parentage}.
This tendency causes incompressibility at the specific values of 
the filling factor $\nu$, as well as the specific properties of 
the elementary excitations of these incompressible GS's
\cite{laughlin,tsui,haldane-hierarchy}.
The avoidance of pair states with small ${\cal R}$ can also be 
mimicked by a composite fermion (CF) transformation\cite{jain} 
in which the ``hard core'' at ${\cal R}<2p+1$ is replaced by 
an attachment of $2p$ vortices or magnetic flux quanta to each 
electron.

Correlations of this type do not generally occur in the excited 
LL's because of different behavior of the electron--electron 
repulsion.
The condition necessary for Laughlin correlations is that the 
interaction pseudopotential\cite{haldane-pseudo}, $V({\cal R})$, 
is super-harmonic at short range\cite{wojs-parentage,wojs-five}, 
and the Coulomb pseudopotential in the $n$th LL, $V_{\rm C}^{(n)}$, 
satisfies this condition only at ${\cal R}\ge2n+1$
\cite{wojs-parentage,wojs-five}.
Consequently, Laughlin correlations in the $n$th LL are not expected 
at $\nu>(2n+2)^{-1}$, and neither will the CF picture be valid at these
fillings.
Therefore, it is not surprising that the half-filled state in the 
$n=1$ LL ($\nu={5\over2}$) is incompressible\cite{willet,moore}, 
even though for $n=0$ all even-denominator fractions are compressible.
It is less obvious that the correlations (and thus the reason for 
incompressibility) at $\nu={7\over3}$ and ${8\over3}$ are different 
from those at $\nu={1\over3}$ and ${2\over3}$, and that the CF model 
does not apply in the $n=1$ LL.

In this note, correlations in the $n\!=\!1$ LL are studied numerically.
The energy spectra and the coefficients of fractional grandparentage 
(CFGP)\cite{wojs-parentage,shalit}, ${\cal G}$, for the lowest energy 
states are calculated.
The pair-correlation functions ${\cal G}({\cal R})$ for the low-energy 
states are analyzed.
The series of Laughlin-correlated states containing electron pairs
is proposed.

In our model\cite{wojs-five}, $N$ electrons are confined on a Haldane 
sphere\cite{haldane-hierarchy}, and the degeneracy of the $n$th LL, 
$g_n=2l_n+1$, is controlled by the strength $2S$ of the magnetic 
monopole inside the sphere ($l_n=S+n$).
The Coulomb matrix elements are calculated assuming zero width of 
the 2DEG, and the inter-LL scattering is neglected.
All lengths and energies are given in the units of $\lambda$ (magnetic 
length) and $e^2/\lambda$.
The many-body states are labeled by the length ($L$) and projection 
($M$) of total angular momentum.

\begin{figure}[t]
\epsfxsize=3.2in
\epsffile{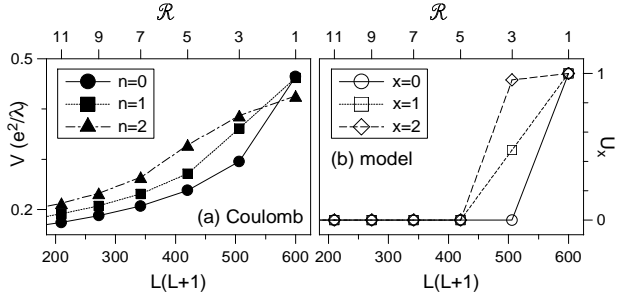}
\caption{
   Coulomb pseudopotentials in different LL's (a) and
   model pseudopotentials $U_x$ (b) 
   calculated on Haldane sphere with $2l=25$.}
\label{fig1}
\end{figure}

On a sphere, ${\cal R}=2l-L$ and the harmonic pseudopotential 
$V_{\rm H}$ is linear in $L(L+1)$\cite{wojs-parentage}.
Only those pseudopotentials $V$ that decrease more quickly than 
$V_{\rm H}$ with increasing ${\cal R}$ cause Laughlin correlations
\cite{wojs-parentage,wojs-five}.
It is clear from Fig.~\ref{fig1}(a) that $V_{\rm C}^{(n)}$, the 
Coulomb pseudopotential in the $n$th LL, is super-harmonic in entire 
range of ${\cal R}$ only for $n\!=\!0$.
To model different behavior of $V_{\rm C}^{(0)}$ and $V_{\rm C}^{(1)}$ 
at short range, a model pseudopotential shown in Fig.~\ref{fig1}(b) 
can be used for which $U_x(1)=1$, $U_x({\cal R}\ge5)=0$, and $U_x(3)=
xV_{\rm H}(3)$, where $V_{\rm H}(3)$ is the ``harmonic'' value such that 
$U_1$ is linear in $L(L+1)$ for ${\cal R}$ between 1 and 5.
While $U_0$ gives similar many-body energy spectra to $V_{\rm C}^{(0)}$, 
the (approximately) harmonic behavior of $V_{\rm C}^{(1)}$ at ${\cal R}
\le5$ is well reproduced by $U_1$.

\begin{figure}[t]
\epsfxsize=3.2in
\epsffile{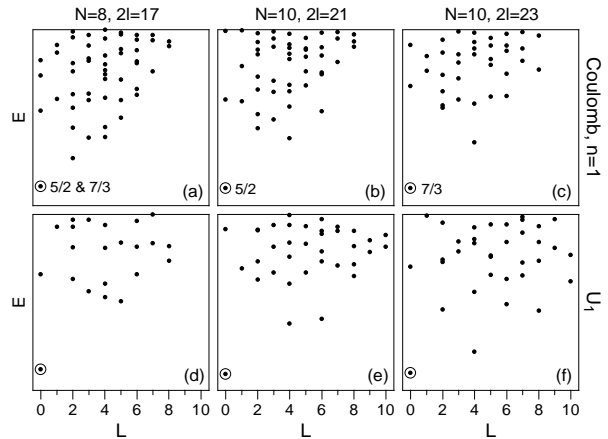}
\caption{
   The $N$-electron energy spectra calculated on Haldane sphere with
   different $2l$ for Coulomb interaction in the $n=1$ LL (abc) and 
   for model interaction $U_1$ (def).}
\label{fig2}
\end{figure}

A few $n=1$ Coulomb energy spectra are compared with the spectra of 
$U_1$ in Fig.~\ref{fig2}.
The circles mark incompressible GS's in each frame, identified as
the $\nu={5\over2}$ (Moore--Read\cite{moore}) and $\nu={7\over3}$ 
states.
The similarity of the corresponding $V_{\rm C}^{(1)}$ and $U_1$ 
spectra (both very different from the $V_{\rm C}^{(0)}$ 
and $U_0$ spectra -- not shown) confirms the fact that the essential 
feature of $V_{\rm C}^{(1)}$ that determines correlations in the 
$n=1$ LL is its harmonic behavior at short range.

\begin{figure}[t]
\epsfxsize=3.2in
\epsffile{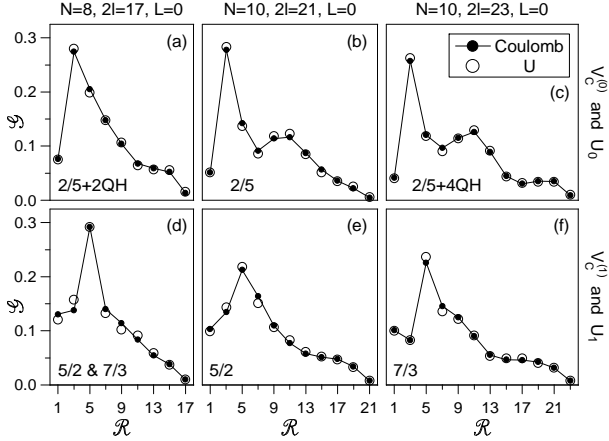}
\caption{
   CFGP profiles of the lowest-energy $N$-electron states at $L=0$,
   calculated on Haldane sphere with different $2l$: 
   (abc) Coulomb interaction in the $n=0$ LL compared to model 
   interaction $U_0$, and
   (def) Coulomb interaction in the $n=1$ LL compared to $U_1$.}
\label{fig3}
\end{figure}

Further confirmation of the essential role of this harmonicity comes 
from the comparison of CFGP profiles (pair-correlation functions) 
${\cal G}({\cal R})$ for the low-energy states, shown in Fig.~\ref{fig3}.
As an example, we display data for the lowest-energy $L=0$ states from 
the spectra of Fig.~\ref{fig2} and the (not shown) analogous spectra 
for $V_{\rm C}^{(0)}$ and $U_0$ (the latter correspond to different 
numbers of quasiholes, QH, in Jain $\nu={2\over5}$ state).
Clearly, the correlations obtained for the Coulomb interaction in 
the $n=0$ and 1 LL's are very different for both one-half and
one-third filling, and they are very well reproduced by the model
interactions $U_0$ and $U_1$, respectively.
The main common feature of ${\cal G}({\cal R})$ for $n=0$ is a strong
minimum at ${\cal R}=1$ that can be viewed as a tendency for the 
electrons to maximally avoid this most strongly repulsive pair state.
Because of the sum rules satisfied by CFGP's\cite{wojs-parentage}: 
(i)  $\sum_{\cal R}{\cal G}({\cal R})=1$ and 
(ii) ${1\over2}N(N-1)\,\sum_{\cal R}L'(L'+1)\,{\cal G}({\cal R})
        =L(L+1)+N(N-2)\,l(l+1)$, 
where $L'=2l-{\cal R}$ and $L$ is the total $N$-electron angular 
momentum, the minimum at ${\cal R}=1$ causes maximum at ${\cal R}=3$.
The harmonicity of $V_{\rm C}^{(1)}$ at $1\le{\cal R}\le5$ results 
in a different ``prescription'' for the CFGP profile that minimizes 
total interaction energy, $E={1\over2}N(N-1)\,\sum_{\cal R}V({\cal R})
\,{\cal G}({\cal R})$, in the $n=1$ LL.
Namely, the total grandparentage from ${\cal R}=1$ and 3 is minimized, 
yielding a maximum at ${\cal R}=5$.

\begin{figure}[t]
\epsfxsize=3.2in
\epsffile{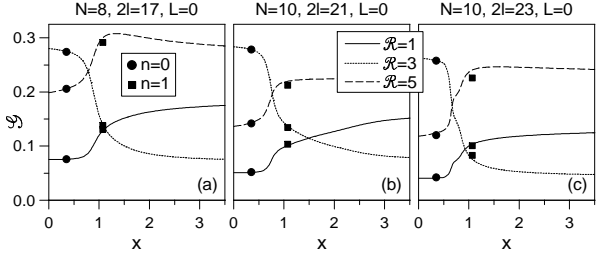}
\caption{
   Dependence of CFGP's ${\cal G}$ from pair states at ${\cal R}
   =1$, 3, and 5, on the anharmonicity parameter $x$ of the model 
   pseudopotential $U_x$, calculated on Haldane sphere with 
   different $2l$ for the lowest-energy $N$-electron states at 
   $L=0$.
   Symbols mark the values of ${\cal G}$ for the Coulomb 
   pseudopotential in the $n=0$ and 1 LL's.}
\label{fig4}
\end{figure}

The optimum pseudopotential $U_x$ to model Coulomb correlations at 
$n\!=\!0$ or 1 can be found from a dependence of the leading CFGP's 
on $x$, shown in Fig.~\ref{fig4} for the same three many-body 
eigenstates of Fig.~\ref{fig3}.
Clearly, the abrupt reconstruction of Laughlin correlations 
characteristic of $V_{\rm C}^{(0)}$ and $U_0$ occurs at $x\approx1$,
and the correlations resulting for $V_{\rm C}^{(1)}$ are best
reproduced by $U_x$ with $x$ near this transition point.
It is noteworthy that ${\cal G}(1)\approx{\cal G}(3)$ at $x\approx1$,
and that the total number of ${\cal R}=1$ pairs, ${1\over2}N(N-1)
{\cal G}(1)$, is roughly equal to ${1\over2}N$.
This supports the idea of electron pairing in the $\nu={5\over2}$ state
\cite{moore}.

\begin{figure}[t]
\epsfxsize=3.2in
\epsffile{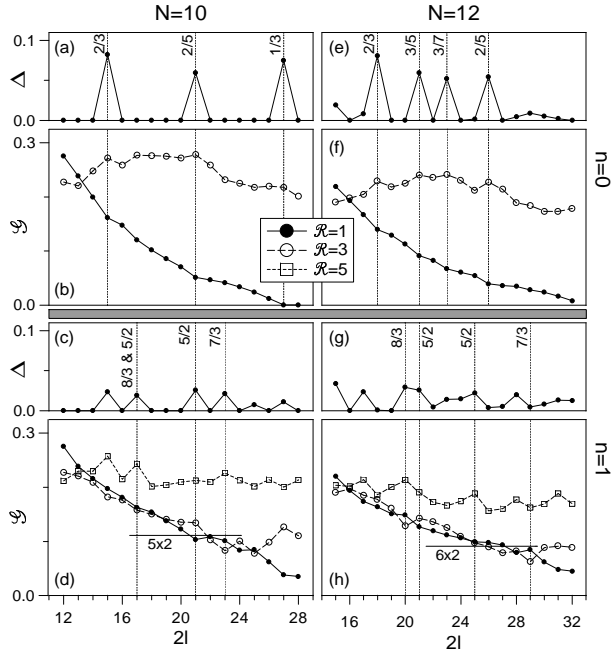}
\caption{
   Dependence of the excitation gap (aceg) and the leading CFGP's 
   (bdfh) on $2l$, calculated on Haldane sphere for $N=10$ (abcd) 
   and 12 (efgh) electrons in the $n=0$ (abef) and $n=1$ (cdgh) LL.}
\label{fig5}
\end{figure}

To identify the finite-size incompressible states in the $n=1$ LL 
(and to rule out the same character of the $\nu={1\over3}$ and 
${7\over3}$ states that coincidentally occur at the same filling 
of the $n\!=\!0$ and 1 LL's in the thermodynamic limit), in Fig.~\ref{fig5} 
we show the dependence of the excitation gap from the $L=0$ GS, $\Delta$, 
and of the leading CFGP's on $2l$, calculated for $N=10$ and 12.
At $n=0$, large gaps $\Delta$ occur only at those $2l$ corresponding 
to Laughlin or Jain GS's at $\nu={2\over3}$, ${2\over5}$, ${1\over3}$, 
etc., and coincide with the downward cusps in ${\cal G}(1)$ and upward 
peaks in ${\cal G}(3)$.
At $n\!=\!1$, the gaps generally occur at different $2l$ than at 
$n\!=\!0$ and coincide with the maxima of ${\cal G}(5)$.
The horizontal lines labeled ``5x2'' and ``6x2'' show the values 
of ${\cal G}(1)=(N-1)^{-1}$ corresponding to the formation of 
${1\over2}N=5$ or 6 pairs with ${\cal R}=1$.

The facts that (i) ${\cal G}(1)\approx(N-1)^{-1}$ over certain range 
of $2l$ for $n=1$ and (ii) Laughlin correlations keeping electrons 
maximally separated from one another no longer occur, suggest that 
electrons may indeed form ${\cal R}=1$ pairs in the $n=1$ LL.
Such pairs would then keep far apart from one another due to the
super-harmonic behavior of $V_{\rm C}^{(1)}$ at larger ${\cal R}$.
Laughlin pair--pair correlations can be formally introduced by
a composite boson (CB) transformation applied to the (bosonic) pairs.
The result is that incompressible Laughlin paired states are expected
at the effective pair filling factors $\nu_2=(2q_2)^{-1}$ with $q_2
=1$, 2, 3, 4, \dots, that translate into the total electron filling 
factors of $\nu=2+2/(q_2+2)={8\over3}$, ${5\over2}$, ${12\over5}$, 
${7\over3}$, \dots\cite{wojs-five}.
On Haldane sphere, these GS's and their particle--hole conjugates are 
expected at $2l={1\over2}N(q_2+2)-(q_2+1)$ and $2l={1\over2}N(q_2+2)+1$, 
respectively.
Remarkably, the latter relation for $q_2=3$ is the same as for the 
Moore--Read (pfaffian) state\cite{moore}, which therefore can be 
interpreted as a Laughlin $\nu_2={1\over6}$ state of ${\cal R}=1$ 
electron pairs.

The incompressible GS's for other values of $q_2$ have not been
confirmed numerically.
However, an $L=0$ GS occurs for any $N$ at $2l=3N-7$ (which gives
$\nu={7\over3}$ in the thermodynamic limit, but is different from 
$2l=3N-3$ of the Laughlin $\nu={1\over3}$ state).
Also, its particle--hole conjugate $\nu={8\over3}$ state occurs
in numerical spectra at any (even) $N$ and $2l={3\over2}N+2$.
Most likely, less than ${1\over2}N$ pairs form in these states, 
and the Laughlin pair--pair and electron--pair correlations occur 
in such two-component plasma of ${\cal R}=1$ pairs and excess 
unpaired electrons (in analogy to the two-component Laughlin 
fluid of charged excitons and electrons\cite{wojs-xminus}).

The authors acknowledge partial support of Grant DE-FG02-97ER45657 
from Materials Science Program -- Basic Energy Sciences of the 
US Dept.\ of Energy.
AW acknowledges support of Polish KBN Grant 2P03B11118.


\vspace*{-0.25in}
\begin{references}
\vspace*{-0.65in}
\footnotesize

\bibitem{prange} 
R. E. Prange and S. M. Girvin, 
   {\sl The Quantum Hall Effect}, 
   Springer-Verlag, New York (1987).

\bibitem{laughlin}
R. Laughlin, 
   Phys. Rev. Lett. {\bf50}, 1395 (1983).

\bibitem{haldane-pseudo} 
F. D. M. Haldane, 
   in Ref.~\cite{prange}, chapter 8, p.~303;
F. D. M. Haldane and E. H. Rezayi,
   Phys. Rev. Lett. {\bf60}, 956 (1988).

\bibitem{wojs-parentage}
A. W\'ojs and J. J. Quinn, 
   Phil. Mag. B {\bf80}, 1405 (2000);
J. J. Quinn and A. W\'ojs, 
   J. Phys.: Cond. Mat. {\bf12}, R265 (2000).

\bibitem{tsui} 
D. C. Tsui, H. L. St\"ormer, and A. C. Gossard,
   Phys. Rev. Lett. {\bf48}, 1559 (1982).

\bibitem{haldane-hierarchy}
F. D. M. Haldane,
   Phys. Rev. Lett. {\bf51}, 605 (1983);
G. Fano, F. Ortolani, and E. Colombo, 
   Phys. Rev. B {\bf34}, 2670 (1986).

\bibitem{jain}
J. K. Jain, 
   Phys. Rev. Lett. {\bf63}, 199 (1989).

\bibitem{wojs-five}
A. W\'ojs, 
   Phys. Rev. B {\bf63}, 125312 (2001).

\bibitem{willet}
R. L. Willet, J. P. Eisenstein, H. L. St\"ormer, D. C. Tsui, 
A. C. Gossard, and  J. H. English,
   Phys. Rev. Lett. {\bf59}, 1776 (1987);
J. P. Eisenstein, H. L. St\"ormer, L. Pfeiffer, and K. W. West,
   Phys. Rev. Lett. {\bf62}, 1540 (1989).

\bibitem{moore}
G. Moore and N. Read,
   Nucl. Phys. B {\bf360}, 362 (1991);
R. H. Morf,
   Phys. Rev. Lett. {\bf80}, 1505 (1998);
E. H. Rezayi and F. D. M. Haldane,
   Phys. Rev. Lett. {\bf84}, 4685 (2000).

\bibitem{shalit}
A. de Shalit and I. Talmi, 
   {\sl Nuclear Shell Theory},
   Academic Press, New York and London (1963).

\bibitem{wojs-xminus}
A. W\'ojs, I. Szlufarska, K. S. Yi, and J. J. Quinn,
   Phys. Rev. B {\bf60}, 11273 (1999).

\end{references}
\end{document}